\documentclass[aps,showpacs,showkeys,preprintnumbers,amsmath,epsf,amssymb,epsfig,nofootinbib]{revtex4}
\usepackage{bm}
\usepackage{graphicx}

\begin{document}

\title{Scale invariant scalar metric fluctuations during inflation: non-perturbative formalism from a 5D vacuum}
\author{$^{1}$ M. Anabitarte  \thanks{%
E-mail address: manabitarte@mdp.edu.ar},$^{1,2}$ M. Bellini
\thanks{E-mail address: mbellini@mdp.edu.ar} and  $^{1,3}$ Jos\'e
Edgar Madriz Aguilar \thanks{E-mail address: madriz@mdp.edu.ar} }
\affiliation{$^{1}$ Departamento de F\'isica, Facultad de Ciencias
Exactas y Naturales, Universidad Nacional de Mar del Plata, Funes
3350, C.P. 7600, Mar del Plata, Argentina. \\ \\ $^{2}$ Consejo
Nacional de Investigaciones Cient\'ificas y T\'ecnicas (CONICET),
Mar del Plata, Argentina.
\\ \\$^{3}$ Departamento de F\'isica, DCI,
Campus Le\'on, Universidad de Guanajuato, C.P. 37150, Le\'on Guanajuato, M\'exico, \\
E-mail: anabitar@mdp.edu.ar; mbellini@mdp.edu.ar;
madriz@mdp.edu.ar}

\begin{abstract}
We extend to 5D an approach of a 4D non-perturbative formalism to
study scalar metric fluctuations of a 5D Riemann-flat de Sitter
background metric. In contrast with the results obtained in 4D,
the spectrum of cosmological scalar metric fluctuations during
inflation can be scale invariant and the background inflaton field
can take sub-Planckian values.
\end{abstract}

 \maketitle

\vskip .5cm


\section{Introduction}

Inflation\cite{Guth} is consistent with current observations of
the temperature anisotropy of the Cosmic Microwave Background
(CMB)\cite{smoot}. Its dynamics may be categorized in two ways:
the original (supercooled) inflation and inflationary models where
dissipation plays an important role. Some examples of these last
are new\cite{Al}, stochastic inflation\cite{si}, warm\cite{berera}
and fresh\cite{fresh} inflation. The most popular model of
supercooled inflation is chaotic inflation\cite{Lin}. In this
model the expansion of the universe is driven by a single scalar
field $\varphi$ called inflaton. At some initial epoch, presumably
the Planck scale, the scalar field is roughly homogeneous and
dominates the energy density. Its initial values are random,
subject to the constraint that energy density is at the Planck
scale. Inflationary model solves several difficulties which arise
from the standard cosmological model, such as the horizon,
flatness, and monopole problems\cite{librolin}. Furthermore, it
provides a mechanism for the creation of primordial density
fluctuations needed to explain the structure formation in the
universe. Inflaton fluctuations are responsible for scalar metric
fluctuations in the universe around the background
Friedmann-Robertson-Walker (FRW) metric:
\begin{displaymath}
dS^2= e^{2\psi}\,dt^2 - a^2(t) e^{-2\psi} \,d\vec{r}^2.
\end{displaymath}
This metric describe non-perturbative gravitational fluctuations
on cosmological scales, in which vector and tensor perturbations
of the metric can be neglected and the fluid can be considered as
irrotational. Furthermore, using the continuity equation on large
scales
\begin{equation}
\frac{\partial\rho}{\partial\tau} = -3{\cal{H}} \left( \rho + P
\right),
\end{equation}
where ${\cal{H}} = {d\over d\tau}\left[{\rm{ln}}\left( a(t)\,
e^{-\psi}\right)\right]$ and $d\tau =e^{\psi} dt$. One can show
that there exists a conserved quantity in time at any order in
perturbation theory
\begin{equation}
f = {\rm{ln}} \left( a e^{-\psi}\right) + \frac{1}{3}
{\Large{\int}}^{\rho} \frac{d\rho'}{\left(P'+\rho'\right)}.
\end{equation}
Considering the invariant $\omega$, which characterizes the
equation of state $P = \omega\,\rho$. The perturbation $\delta
f=-\psi +{1\over 3(1+\omega)}
{\rm{ln}}\left(\rho/\bar\rho\right)$, is a gauge-invariant
quantity representing the non-linear extension of the curvature
perturbation for adiabatic fluids on uniform energy density
hypersurfaces on superhorizon scales\cite{kolb,ab}.

In this letter we examine scalar metric fluctuations from a 5D
vacuum state, which is defined on a 5D background Riemann-flat
metric, using ideas of Modern Kaluza-Klein theory. This theory
allows the fifth coordinate to play an important physical role. In
this framework, the Induced Matter (or Space-Time-Matter)
theory\cite{wbook}, all classical physical quantities, such as
matter density and pressure, are susceptible of a geometrical
interpretation. The mathematical basis of it is the Campbell's
theorem\cite{campbell}, which ensures an embedding of 4D general
relativity with sources in a 5D theory whose field equations are
apparently empty. That is, the Einstein equations
$G_{\alpha\beta}=-8\pi G\,T_{\alpha\beta}$ (we use $c=\hbar=1$
units), are embedded perfectly in the Ricci-flat equations
$R_{AB}=0$. Other version of 5D gravity, which is mathematically
similar, is the membrane theory, in which gravity propagates
freely on the 5D bulk and the interactions of particles are
confined to a 4D hypersurface called "brane"\cite{rs}. Both
versions of 5D general relativity are in agreement with
observations.

\section{Scalar metric fluctuations on a de Sitter spacetime}

In this section we study the scalar metric fluctuations of a 5D
spacetime background metric $({\cal M}, g)$, which is
Riemann-flat. Physically, the background metric here employed
describes a 5D extension of the usual de Sitter spacetime.
Similarly to the Induced Matter theory, here the flatness of the
5D background metric defines a 5D apparent vacuum. In the first
part of the present section, we obtain the corresponding 5D
Einstein field equations. In the second part, we derive the 4D
induced dynamics for the scalar metric fluctuations.

\subsection{The non-perturbative gauge-invariant scalar metric fluctuations}

In order to study scalar metric fluctuations on a 5D de Sitter
spacetime, we shall use the background line element\cite{lebe}
\begin{equation}\label{s1}
dS^2_b=l^2 dN^2-l^2 e^{2N} dr^2 -dl^2,
\end{equation}
where $dr^2=\delta_{ij}dx^{i}dx^{j}$, $x^i$ are the 3D cartesian
space-like dimensionless coordinates, $N$ is a dimensionless
time-like coordinate and $l$ is the space-like non-compact extra
coordinate, which has length units. The non-perturbative metric
fluctuations of the background metric (\ref{s1}), are introduced
in our analysis by the line element
\begin{equation}\label{s2}
dS^2 =  l^2 e^{2\psi} dN^2 - l^2 e^{2(N- \psi)}dr^2-dl^2,
\end{equation}
where the metric function $\psi(N,\vec{r},l)$ describes the gauge-invariant metric fluctuations.\\

Now let us consider a non-massive test scalar field
$\varphi=\varphi (x^{\alpha},l)$ defined on the 5D spacetime
(\ref{s2}). The dynamics of $\varphi$ can be derived from the
action
\begin{equation}\label{s3}
^{(5)}{\cal S}={\int} d^4 x \    dl \sqrt{\left|\frac{^{(5)}
 g}{^{(5)} g_0}\right|} \left(
\frac{^{(5)} R}{16\pi G}+ \frac{1}{2}g^{AB} \varphi_{,A}
\varphi_{,B} \right),
\end{equation}
where $^{(5)}R$ is the 5D Ricci scalar,  $^{(5)}g$ is the
determinant of the metric (\ref{s2}) which for the case of the
background metric (\ref{s1}) results in $^{(5)}\bar g =
l^{8}\exp(6N)$ and for the perturbed metric (\ref{s2}) becomes
$^{(5)}g = l^{8}\exp[6(N-4\psi)]$. Furthermore $^{(5)}g_0$ is a
dimensionalization constant. The scalar field $\varphi$ is a
purely kinetic scalar field without
a potential which means that it is non-massive and therefore there are no
interactions of this field with its environment.\\

In order to describe the scalar metric fluctuations consistently
with the idea of a 5D apparent vacuum, we will require the
perturbed metric (\ref{s2}) to be Ricci-flat i.e $^{(5)}R_{AB}=0$
for (\ref{s2}). This condition is given explicitly by
\begin{eqnarray}
^{(5)} R & = & \frac{2e^{2(\psi-N)}}{l^2} \left\{\nabla_{r}^2\psi-(\nabla_{r}\psi)^2
+ e^{-2(\psi-N)}\left[e^{-2\psi}\left(15\frac{\partial\psi}{\partial N}-9\left(\frac{\partial\psi}{\partial N}\right)^2
-6\right)- \right.\right.\nonumber\\
&&\left.\left.-2l^2\left(\frac{\partial^2\psi}{\partial
l^2}-2\left(\frac{\partial\psi}{\partial l}\right)^{2}\right)-10
l\frac{\partial\psi}{\partial l}+6 \right] \right\}=0.\label{s4}
\end{eqnarray}
From the action (\ref{s3}), the field equation for the test scalar
field $\varphi$ on the perturbed metric (\ref{s2}) reads
\begin{equation}\label{s5}
\frac{\partial^2\varphi}{\partial
N^2}+3\frac{\partial\varphi}{\partial
N}-4\frac{\partial\psi}{\partial N}\frac{\partial\varphi}{\partial
N}-e^{4\psi-2N}\nabla_{r}^{2}\varphi
-l^{2}e^{2\psi}\left[\frac{\partial ^{2}\varphi}{\partial
l^2}+2\left(\frac{2}{l}-\frac{\partial\psi}{\partial
l}\right)\frac{\partial\varphi}{\partial l}\right]=0.
\end{equation}
The diagonal components of the 5D field equations $G_{AB}=8\pi G
T_{AB}$, are given by
\begin{eqnarray}
&&3e^{2\psi}+6\frac{\partial\psi}{\partial N}-3\left(\frac{\partial\psi}{\partial N}\right)^{2}
-3+e^{4\psi-2N}\left[(\vec{\nabla}_{r}\psi)^{2}-2\nabla^{2}_{r}\psi\right]+e^{2\psi}
\left[6l^{2}\left(\frac{\partial\psi}{\partial l}\right)^{2}-12 l\frac{\partial\psi}{\partial l}-3l^{2}
\frac{\partial^{2}\psi}{\partial l^2}\right]\nonumber\\
\label{s6} &&=8\pi G
e^{2(\psi-N)}\left[\frac{1}{2}\left(\frac{\partial\varphi}{\partial
N}\right)^{2}+\frac{1}{2}e^{4\psi-2N}(\vec{\nabla}_{r}\varphi)^{2}+\frac{l^2}{2}e^{2\psi}\left(\frac{\partial\varphi}{\partial
l}\right)^{2}\right],
\end{eqnarray}

\begin{eqnarray}
&& 24\frac{\partial\psi}{\partial
N}-15\left(\frac{\partial\psi}{\partial
N}\right)^{2}+6\frac{\partial^{2}\psi}{\partial
N^2}+9\left(e^{2\psi}-1\right)+e^{4\psi-2N}(\vec{\nabla}_{r}\psi)^{2}+e^{2\psi}\left[6l^{2}\left(\frac{\partial\psi}{\partial
l}\right)^{2}-12l\frac{\partial\psi}{\partial
l}-3l^{2}\frac{\partial^{2}\psi}{\partial l^2}\right]
\nonumber\\
\label{s7} && =-8\pi G
e^{4\psi-2N}\left[-\frac{1}{2}(\vec{\nabla}_{r}\varphi)^{2}+\frac{3}{2}e^{-4\psi}e^{2N}\left(\frac{\partial\varphi}{\partial
N}\right)^{2}
-\frac{3}{2}l^{2}e^{-2\psi}e^{2N}\left(\frac{\partial\varphi}{\partial
l}\right)^{2}\right],
\end{eqnarray}

\begin{eqnarray}
&& e^{4\psi-2N}\nabla_{r}^{2}\psi -15\frac{\partial\psi}{\partial N}
+9\left(\frac{\partial\psi}{\partial N}\right)^{2}+6-3\frac{\partial^{2}\psi}{\partial N^2}
-e^{4\psi-2N}(\vec{\nabla}_{r}\psi)^{2}-6e^{2\psi}+6le^{2\psi}\frac{\partial\psi}{\partial l}\nonumber\\
\label{s8} && = 8\pi G
l^{2}e^{2\psi}\left[\frac{1}{2}\left(\frac{\partial\varphi}{\partial
l}\right)^{2}+\frac{1}{2}l^{-2}e^{-2\psi}\left(\frac{\partial\varphi}{\partial
N}\right)^{2}-\frac{1}{2}l^{-2}e^{2\psi}e^{-2N}(\vec{\nabla}_{r}\varphi)^{2}\right],
\end{eqnarray}
while the non-diagonal ones are
\begin{eqnarray}\label{s9}
\frac{\partial^{2}\psi}{\partial x^{i}\partial
N}-\frac{\partial\psi}{\partial N}\frac{\partial\psi}{\partial
x^i}+\frac{\partial\psi}{\partial x^i}&=& 4\pi G
\frac{\partial\varphi}{\partial N}\frac{\partial\varphi}{\partial
x^i},
\\
\label{s10}
 \frac{\partial^{2}\psi}{\partial l\partial N}-2\frac{\partial\psi}{\partial N}\frac{\partial\psi}{\partial l}
 +2\frac{\partial\psi}{\partial l}&=& \frac{8}{3} \pi G \:\frac{\partial \varphi}{\partial N} \:\frac{\partial \varphi}{\partial l},\\
\label{s11}
\frac{\partial ^2\psi}{\partial x^{i}\partial l}-2\frac{\partial\psi}{\partial x^{i}}\frac{\partial\psi}{\partial l}
&=& 8 \pi G \frac{\partial \varphi}{\partial x^i} \frac{\partial \varphi}{\partial l}\\
\label{s12} 2\frac{\partial\psi}{\partial x^{i}}\:\frac{\partial
\psi}{\partial x^j} &=& -8\pi G\: \frac{\partial\varphi}{\partial
x^i}\:\frac{\partial\varphi}{\partial x^j},\qquad \mbox{for}\qquad
i\ne j.
\end{eqnarray}
By using the expression (\ref{s4}), the linear combination
[$3\times \,eq(\ref{s6})+eq(\ref{s7})+2\times \,eq(\ref{s8})$]
yields
\begin{eqnarray}
&&\frac{\partial^{2}\psi}{\partial N^2}+7\frac{\partial\psi}{\partial N}
-4\left(\frac{\partial\psi}{\partial N}\right)^{2}+3\left(e^{2\psi}-1\right)+\frac{2}{3}
e^{4\psi-2N}(\vec{\nabla}_{r}\psi)^{2}-e^{4\psi-2N}\nabla^{2}_{r}\psi +e^{2\psi}\left[\frac{16}{3}l^{2}
\left(\frac{\partial\psi}{\partial l}\right)^{2}-\frac{28}{3}l\frac{\partial\psi}{\partial l}
-\frac{8}{3}l^{2}\frac{\partial^{2}\psi}{\partial l^2}\right]\nonumber\\
\label{s13} &&= -\frac{4\pi
G}{3}\left[\left(\frac{\partial\varphi}{\partial
N}\right)^{2}+e^{4\psi-2N}(\vec{\nabla}_{r}\varphi)^{2}+l^{2}e^{2\psi}\left(\frac{\partial\varphi}{\partial
l}\right)^{2}\right].
\end{eqnarray}
These expressions provide us with the 5D dynamics of the scalar
metric fluctuations $\psi(t,\vec{r},l)$. However, our main
interest, more than describing the 5D dynamics, consists in
describing the 4D dynamics induced from these perturbed field
equations.

\subsection{The induced 4D dynamics for the non-perturbative gauge-invariant scalar metric fluctuations}

The field equations obtained in the preceding section can be used
to derive an induced 4D dynamics for the 4D scalar fluctuations.
To go down from 5D to 4D we assume that our 5D spacetime can be
foliated by the family of hypersurfaces $\Sigma _H:l=l_0=H^{-1}$,
$H$ being the Hubble constant. Physically, the class of observers
living on a generic $\Sigma _H$ are those whose 5-velocity along
the fifth coordinate is null: $U^{l}=0$. Thus, after passing to a
more physical coordinates on $\Sigma _H$ through the coordinate
transformation $t=N/H$, $R=r/H$, the background metric (\ref{s1})
yields the background-induced metric
\begin{equation}\label{w1}
ds^2_b = dt^2 - e^{2H t} dR^2,
\end{equation}
where $t$ is the cosmic time. This metric describes the de Sitter
expansion of the early universe characterized by an equation of
state $p_{b}=-\rho _{b}=-3H^{2}/(8\pi G)$, $p_b$ and $\rho_b$
being the background pressure and energy densities. Similarly, the
perturbed metric (\ref{s2}) induces on $\Sigma _H$ the metric
\begin{equation}\label{w2}
ds^2 = e^{2\Psi} dt^2 - e^{2 \left(H t-\Psi\right)} dR^2,
\end{equation}
where $\Psi(t,\vec{R})\equiv\psi(t,\vec{R},l)|_{l=H^{-1}} $
describes the scalar metric fluctuations induced on $\Sigma _H$.
The 4D Ricci scalar for the perturbed metric (\ref{w2}) is given
by
\begin{equation}\label{w3}
^{(4)} R = 2 \left[6H^2+9\left(\frac{\partial\Psi}{\partial
t}\right)^{2}-3\frac{\partial^{2}\Psi}{\partial
t^2}-15H\frac{\partial\Psi}{\partial t}+
e^{4\Psi-2Ht}\left(\nabla_{R}^{2}\Psi-(\vec{\nabla}_{R}\Psi)^{2}\right)\right]
e^{-2\Psi}.
\end{equation}
Note that in the absence of scalar metric fluctuations we recover the value of the Ricci scalar for the
background-induced metric (\ref{w1}), $^{(4)}R=12H^2$.\\

The diagonal Einstein field equations induced from the 5D field
equations (\ref{s6}), (\ref{s7}) and (\ref{s8}), on the
hypersurface $\Sigma _H$, read
\begin{eqnarray}\label{w4}
&&3H^{2}e^{2\Psi}+6H\frac{\partial\Psi}{\partial t}-3\left(\frac{\partial\Psi}{\partial t}
\right)^{2}-3H^{2}+e^{4\Psi-2Ht}\left[(\vec{\nabla}_{R}\Psi)^{2}-2\nabla_{R}^{2}\Psi\right]
+H^{2}e^{2\Psi}\left[6l^{2}\left(\frac{\partial\psi}{\partial l}\right)^{2}-12l\frac{\partial\psi}{
\partial l}-3l^{2}\frac{\partial^{2}\psi}{\partial l^2}\right]_{l=H^{-1}}   \nonumber\\
&&=  4 \pi G
e^{2(\Psi-Ht)}\left[\left(\frac{\partial\Phi}{\partial
t}\right)^{2}+e^{4\Psi-2Ht}(\vec{\nabla}_{R}\Phi)^{2}+H^{2}e^{2\Psi}\left(l\frac{\partial\varphi}{\partial
l}\right)^{2}_{l=H^{-1}} \right],
\end{eqnarray}

\begin{eqnarray}\label{w5}
&& 24H\frac{\partial\Psi}{\partial t}-15\left(\frac{\partial\Psi}{\partial t}\right)^{2}+6\frac{
\partial^{2}\Psi}{\partial t^2}+9H^{2}\left(e^{2\Psi}-1\right)+e^{4\Psi-2Ht}(\vec{\nabla}_{R}\Psi)^{2}
+H^{2}e^{2\Psi}\left[6l^{2}\left(\frac{\partial\psi}{\partial l}\right)^{2}-12l\frac{\partial\psi}{\partial l}
-3l^{2}\frac{\partial^{2}\psi}{\partial l^2}\right]_{l=H^{-1}}\nonumber \\
&& = -4 \pi G \left[ 3\left(\frac{\partial\Phi}{\partial
t}\right)^{2}-e^{4\Psi-2Ht}(\vec{\nabla}_{R}\Phi)^{2}-3H^{2}e^{2\Psi}\left(l\frac{\partial\varphi}{\partial
l}\right)^{2}_{l=H^{-1}}\right],
\end{eqnarray}

\begin{eqnarray}\label{w6}
&& e^{4\Psi-2Ht}\nabla_{R}^{2}\Psi-15H\frac{\partial\Psi}{\partial
t}+9\left(\frac{\partial\Psi}{\partial
t}\right)^{2}+6H^{2}-3\frac{\partial^{2}\Psi}{\partial
t^2}-e^{4\Psi-2Ht}(\vec{\nabla}_{R}\Psi)^{2}-6H^{2}e^{2\Psi}\left(l\frac{\partial\psi}{\partial
l}\right)_{l=H^{-1}}
\nonumber \\
&& = 4 \pi G \left[\left(\frac{\partial\Phi}{\partial
t}\right)^{2}-e^{4\Psi-2Ht}(\vec{\nabla}_{R}\Phi)^{2}+H^{2}e^{2\Psi}\left(l\frac{\partial\varphi}{\partial
l}\right)^{2}_{l=H^{-1}} \right],
\end{eqnarray}
where $\Phi(t,\vec{r})\equiv\varphi(t,\vec{r},l)|_{l=H^{-1}}$ is
the massive scalar field induced on $\Sigma _H$. The non-diagonal
induced 4D Einstein equations can be derived by evaluating the
equations (\ref{s9}), (\ref{s10}), (\ref{s11}) and (\ref{s12}) on
$\Sigma _{H}$.\\

The induced dynamics of $\Phi(t,\vec{r})$ [see Eq. (\ref{s5})] on
$\Sigma _{H}$, is
\begin{equation}\label{w7} \frac{\partial^{2}\Phi}{\partial
t^2}+3H\frac{\partial\Phi}{\partial
t}-4\frac{\partial\Psi}{\partial t}\frac{\partial\Phi}{\partial
t}-e^{4\Psi-2Ht}\nabla_{R}^{2}\Phi -
H^{2}l^{2}e^{2\Psi}\left[\frac{\partial^{2}\varphi}{\partial
l^{2}}+2\left(\frac{2}{l}-\frac{\partial\psi}{\partial
l}\right)\frac{\partial\varphi}{\partial l}\right]_{l=H^{-1}}=0,
\end{equation}
whereas the dynamics of the induced scalar metric fluctuations
$\Psi$, is given by the expression
\begin{eqnarray}
&&\frac{\partial^{2}\Psi}{\partial t^2}+7H\frac{\partial\Psi}{\partial t}
-4\left(\frac{\partial\Psi}{\partial t}\right)^{2}+3H^{2}\left(e^{2\Psi}-1\right)
+\frac{2}{3}e^{4\Psi-2Ht}(\vec{\nabla}_{R}\Psi)^{2}-e^{4\Psi-2Ht}\nabla_{R}^{2}\Psi
+ \frac{H^{2}e^{2\Psi}}{3}\left[16l^{2}\left(\frac{\partial\psi}{\partial l}\right)^{2}\right.\nonumber\\
\label{w8} &&\left.-28l\frac{\partial\psi}{\partial
l}-8l^{2}\frac{\partial^{2}\psi}{\partial l^2}\right]_{l=H^{-1}}
=-\frac{4\pi G}{3}\left[\left(\frac{\partial\Phi}{\partial
t}\right)^{2}+e^{4\Psi-2Ht}(\vec{\nabla}_{R}\Phi)^{2}+H^{2}e^{2\Psi}\left(l\frac{\partial\varphi}{\partial
l}\right)^{2}_{l=H^{-1}}\right],
\end{eqnarray}
which is obtained by evaluating (\ref{s13}) on $\Sigma _H$.\\

The effective 4D action on $\Sigma _H$ derived from the 5D action
(\ref{s3}), reads
\begin{equation}\label{w8}
^{(4)}{\cal S}_{eff}=\int d^{4}x\,\sqrt{\left|\frac{^{(4)}
 g}{^{(4)} g_0}\right|} \left( \frac{^{(4)} R}{16\pi G}+
\frac{1}{2}g^{\mu\nu} \Phi_{,\mu} \Phi_{,\nu} + V \right),
\end{equation}
where $^{(4)}g$ is the determinant of the 4D induced metric, which
in the case of background metric (\ref{w1}) yields
$^{(4)}g=\exp(6Ht)$. Moreover, in the case of the perturbed metric
(\ref{w2}) gives $^{(4)}g=\exp[2(3Ht-2\Psi)]$, and $^{(4)}g_0$ is
a dimensionless constant. The induced 4D effective potential $V$
in the action (\ref{w8}), has the form
\begin{equation}\label{w9}
V = -\left.\frac{1}{2} g^{ll}
\left(\frac{\partial\varphi}{\partial l}\right)^2\right|_{N=H
t,R=r/H,l=l_0=1/H}\,.
\end{equation}
Given the quantum nature of the fields $\Phi$ and $\Psi$ it seems
convenient to use the quantization procedure. To do it, we impose
the commutation relations
\begin{equation}\label{w10}
\left[\Phi(t,\vec R), \Pi^0_{(\Phi)}(t,\vec R')\right] = i\,
\delta^{(3)}\left(\vec R-\vec R'\right),\quad \left[\Psi(t,\vec
R), \Pi^0_{(\Psi)}(t,\vec R')\right] = i\, \delta^{(3)}\left(\vec
R-\vec R'\right),
\end{equation}
from which we can derive the commutation relations
\begin{equation}\label{w11}
\left[\Phi(t,\vec R), \dot\Phi(t,\vec R')\right] = i e^{2\Psi}
\sqrt{\left|\frac{^{(4)} g_0}{^{(4)} g}\right|}
\delta^{(3)}\left(\vec R-\vec R'\right),\quad \left[\Psi(t,\vec
R), \dot\Psi(t,\vec R')\right] = i \frac{4}{9} \pi G
e^{2\Psi}\sqrt{\left|\frac{^{(4)} g_0}{^{(4)} g}\right|}
\delta^{(3)}\left(\vec R-\vec R'\right),
\end{equation}
with $|\,^{(4)}g|=[\exp(3Ht-2\Psi)/H^{3}]^{2}$ being the
determinant of the effective perturbed metric (\ref{w2}), and $\Pi
^{0}_{(\Psi)}=\dot{\Phi}\exp(-2\Psi)\sqrt{|\,^{(4)}g/^{(4)}g_{0}|}$
and $\Pi^{0}_{(\Phi)}=[6/16\pi
G](6\dot{\Psi}-5H)\exp(-2\Psi)\sqrt{|\,^{(4)}g/^{(4)}g_{0}|}$ the
momentums conjugate to $\Phi$ and $\Psi$, respectively.

\subsection{5D scalar metric fluctuations in the weak field limit}

In the previous sections we have derived, in a general manner, the
5D and 4D dynamical equations for the inflaton field and
gauge-invariant metric fluctuations on a perturbed de Sitter
spacetime, regarding that these fluctuations are non-perturbative
in nature. Until this moment, we have not imposed any restriction
about the amplitude of these scalar fluctuations. However, as it
is well known these fluctuations are very small on cosmological
scales (in particular in inflationary scenarios). Thus, a
first-order approximation in the gauge scalar fluctuations of the
form $e^{\pm 2\psi}\simeq 1\pm 2\psi$, will be sufficient in order
to have a good description of these
fluctuations during inflation, in the present formalism. \\

Under this weak field limit approximation, linearizing the
equation (\ref{s5}) with respect to $\psi$, one obtains
\begin{equation}\label{x1}
\frac{\partial^{2}\varphi}{\partial
N^2}+3\frac{\partial\varphi}{\partial
N}-e^{-2N}\nabla_{r}^{2}\varphi-4\frac{\partial\psi}{\partial
N}\frac{\partial\varphi}{\partial
N}-l^{2}\left[\frac{\partial^{2}\varphi}{\partial
l^2}+\frac{4}{l}\frac{\partial\varphi}{\partial
l}-2\frac{\partial\psi}{\partial l}\frac{\partial\varphi}{\partial
l}\right]-2l^{2}\psi\left[\frac{\partial^{2}\varphi}{\partial
l^2}+\frac{4}{l}\frac{\partial\varphi}{\partial l}\right]=0.
\end{equation}
Performing a semiclassical approximation for the 5D scalar field
$\varphi$ in the form
$\varphi(N,\vec{r},l)=\varphi_{b}(N,l)+\delta\varphi
(N,\vec{r},l)$ (with $\varphi _{b}$ denoting the background part
of $\varphi$ and $\delta\varphi$ denoting the quantum fluctuations
of $\varphi$), Eq. (\ref{x1}) results in the system
\begin{eqnarray}\label{x2}
&&\frac{\partial^{2}\varphi_{b}}{\partial N^2}+3\frac{\partial\varphi_{b}}{\partial N}
-\left[l^{2}\frac{\partial^{2}\varphi_b}{\partial l^2}+4l\frac{\partial\varphi _{b}}{\partial l}\right]=0,\\
\label{x3} && \frac{\partial^{2}\delta\varphi}{\partial
N^2}+3\frac{\partial\delta\varphi}{\partial
N}-e^{-2N}\nabla_{r}^{2}\delta\varphi-\left[l^{2}\frac{\partial^{2}\delta\varphi}{\partial
l^2}+4l\frac{\partial\delta\varphi}{\partial
l}\right]-2\psi\left[l^{2}\frac{\partial^{2}\varphi_{b}}{\partial
l^2} + 4l\frac{\partial\varphi_{b}}{\partial l}\right]=0.
\end{eqnarray}
The expression (\ref{x2}) gives the dynamics of the background
scalar field $\varphi(N,\vec{r},l)$, whereas Eq. (\ref{x3})
describes the dynamics for the quantum fluctuations
$\delta\varphi(N,\vec{r},l)$ in terms of the scalar metric
fluctuations $\psi$ and the background field $\varphi _{b}$.
Linearizing the 5D field equations (\ref{s6}), (\ref{s7}) and
(\ref{s8}), we obtain the independent Einstein equations for the
background field $\varphi _b$, on (\ref{s1})
\begin{eqnarray}\label{x4}
\left(\frac{\partial\varphi_b}{\partial N}\right)^2 - l^2 \left(\frac{\partial\varphi_b}{\partial l}\right)^2 &=&0,\\
\label{x6} \left(\frac{\partial\varphi_b}{\partial N}\right)^2 +
l^2 \left(\frac{\partial\varphi_b}{\partial l}\right)^2 &=&0,
\end{eqnarray}
and for the field $\delta\varphi$, we have
\begin{eqnarray}\label{x7}
&& 6\psi+6\frac{\partial\psi}{\partial
N}-2e^{-2N}\nabla_{r}^{2}\psi-\left[12l\frac{\partial\psi}{\partial
l}+3l^{2}\frac{\partial^{2}\psi}{\partial l^2}\right] = 8\pi
G\left[\frac{\partial\varphi_b}{\partial N}\frac{\partial
\delta\varphi}{\partial
N}+l^{2}\frac{\partial\varphi_{b}}{\partial
l}\frac{\partial\delta\varphi}{\partial
l}+3l^{2}\psi\left(\frac{\partial\varphi_{b}}{\partial
l}\right)^{2}+2\psi\left(\frac{\partial\varphi_{b}}{\partial
N}\right)^{2}\right]
\\
\label{x8}
&& 8\frac{\partial\psi}{\partial N}+2\frac{\partial ^{2}\psi}{\partial N^2}+6\psi-
\left[4\frac{\partial\psi}{\partial l}+l^{2}\frac{\partial ^{2}\psi}{\partial l^2}\right]
= -8\pi G\left[\frac{\partial\varphi _b}{\partial N}\frac{\partial\delta\varphi}{\partial N}
-l^{2}\frac{\partial\varphi_{b}}{\partial l}\frac{\partial\delta\varphi}{\partial l} + l^{2}\psi
\left(\frac{\partial\varphi_{b}}{\partial l}\right)^{2}-2l^{2}\psi\left(\frac{\partial\varphi_b}{\partial l}\right)^{2}\right] \\
\label{x9} && e^{-2N}\nabla_{r}^{2}\psi -
15\frac{\partial\psi}{\partial
N}-3\frac{\partial^{2}\psi}{\partial
N^2}-12\psi+6l\frac{\partial\psi}{\partial l} = 8\pi G
\left[l^{2}\frac{\partial\varphi_b}{\partial
l}\frac{\partial\delta\varphi}{\partial l}+\frac{\partial\varphi
_b}{\partial N}\frac{\partial\delta\varphi}{\partial
N}+l^{2}\psi\left(\frac{\partial\varphi_b}{\partial
l}\right)^{2}\right].
\end{eqnarray}
Now, linearizing the non-diagonal field equations (\ref{s9}),
(\ref{s10}) and (\ref{s11}), we obtain the system
\begin{eqnarray}\label{x10}
&& \frac{\partial^{2}\psi}{\partial x^{i}\partial N}+\frac{\partial\psi}{\partial x^i}=
4\pi G\frac{\partial\varphi_b}{\partial N}\frac{\partial\delta\varphi}{\partial x^i},\\
\label{x11}
&& \frac{\partial^{2}\psi}{\partial l\partial N}+2\frac{\partial\psi}{\partial l}=\frac{8\pi G}{3}
\left(\frac{\partial\delta\varphi}{\partial N}\frac{\partial\varphi_b}{\partial l}+\frac{\partial\varphi_b}{\partial N}
\frac{\partial\delta\varphi}{\partial l}\right),\\
\label{x14} && \frac{\partial ^{2}\psi}{\partial x^{i}\partial
l}=8\pi G\left(\frac{\partial\varphi_b}{\partial
x^i}\frac{\partial\delta\varphi}{\partial
l}+\frac{\partial\delta\varphi}{\partial
x^i}\frac{\partial\varphi_b}{\partial l}\right).
\end{eqnarray}
By using (\ref{x6}), the linearized equation (\ref{s13}) yields
\begin{equation}\label{x15}
\frac{\partial^{2}\psi}{\partial
N^2}+7\frac{\partial\psi}{\partial N} +
6\psi-e^{-2N}\nabla_{r}^{2}\psi-\frac{2}{3}\left[14l\frac{\partial\psi}{\partial
l}+4l^{2}\frac{\partial^{2}\psi}{\partial l^2}\right] =
-\frac{8\pi G}{3}\left[\frac{\partial\varphi_b}{\partial
N}\frac{\partial\delta\varphi}{\partial N} + \psi
l^{2}\left(\frac{\partial\varphi _b}{\partial l}\right)^{2}+
l^{2}\frac{\partial\varphi_{b}}{\partial
l}\frac{\partial\delta\varphi}{\partial l}\right].
\end{equation}
This equation describes the dynamics of the 5D metric fluctuations
$\psi$ in terms of the 5D inflaton field fluctuations
$\delta\varphi$, at first order in $\psi$ and $\delta\varphi$.
Note that this equation also can be obtained by using a linear
combination of Eqs. (\ref{x7}), (\ref{x8}) and (\ref{x9}).

\subsection{Induced 4D dynamics for metric fluctuations in the weak field limit}

Inherent to the semiclassical approximation for
$\varphi(N,\vec{r},l)$, is the semiclassical approximation for the
induced 4D inflaton field $\Phi(t,\vec{r})$. This means that the
semiclassical approximation $\Phi(t,\vec{r})=\Phi_{b}(t)
+\delta\Phi(t,\vec{r})$ is also valid, where $\Phi_{b}(t)$ is the
4D induced background inflaton field. Evaluating the expression
(\ref{x2}) on $\Sigma _0$, we obtain
\begin{equation}\label{af1}
\frac{\partial^{2}\Phi _b}{\partial
t^2}+3H\frac{\partial\Phi_b}{\partial t}+H^{2}m^{2}\Phi_{b}=0
\end{equation}
where we have used $[l^{2}(\partial^{2}\varphi_b/\partial
l^2)+4l(\partial\varphi_b/\partial l)]|_{l=H^{-1}}=-m^{2}\Phi_b$,
$m$ being a dimensionless separation constant. In addition,
according to the action (\ref{w8}) and the induced metric
(\ref{w1}), the background field $\Phi_b$ satisfies the dynamical
Friedmann equation
\begin{equation}\label{af2}
\left(\frac{\partial\varphi_b}{\partial
t}\right)^{2}+H^{2}\left(l\frac{\partial\varphi_b}{\partial
l}\right)^{2}_{l=H^{-1}}=\frac{3H^{2}}{4\pi G}.
\end{equation}
A particular solution of (\ref{af1}), which is valid at the
beginning of inflation (when slow rolling conditions are
fulfilled\cite{cop}), is given by $(\partial\Phi_b/\partial t)=0$.
In this case necessarily $m=0$ and Eq. (\ref{af2}) leads to a
constant solution for the induced field $\Phi _b$
\begin{equation}\label{m2}
\Phi_{b}=\frac{1}{\sqrt{12\pi G}}=\frac{M_{p}}{\sqrt{12\pi}},
\end{equation}
where we have used that $[l^{2}(\partial\varphi_{b}/\partial
l)^{2}]|_{l=H^{-1}}=9\Phi_{b}^{2}$. The remarkable in this result
is the fact that in this formalism $\Phi _b$ remains below
the Planckian mass. \\

Now according to Eqs. (\ref{x11}) to (\ref{x15}), we obtain that
on the hypersurface $\Sigma _H$ the scalar fluctuations $\Psi$
satisfy
\begin{equation}\label{m3}
\frac{\partial^{2}\Psi}{\partial
t^2}+7H\frac{\partial\Psi}{\partial
t}-e^{-2Ht}\nabla_{R}^{2}\Psi+(7H^2-\lambda^{2})\Psi=0,
\end{equation}
where we have used $(8\pi
G)H^2[l^{2}(d^{2}\psi/dl^2)+4l(d\psi/dl)]|_{l=H^{-1}} = \lambda^2
\Psi$, with the separation constant $\lambda$  having mass units.
Equation (\ref{x10}), evaluated on $\Sigma _H$, gives the
condition $(\partial\Psi/\partial t)=-H\Psi$. Thus, by making use
of this condition, Eq. (\ref{m3}) can be written as
\begin{equation}\label{md2}
\frac{\partial^{2}\Psi}{\partial
t^2}+3H\frac{\partial\Psi}{\partial
t}-e^{-2Ht}\nabla_{R}^{2}\Psi+(3H^{2}-\lambda^2)\Psi=0.
\end{equation}
Following a canonical quantization process, the field $\Psi$ can
be written as a Fourier expansion
\begin{equation}\label{m4}
\Psi(t,\vec{R}) = \frac{1}{(2\pi)^{3/2}}\int d^{3}k
\left[a_{k}\,e^{i\vec{k}
\cdot\vec{R}}\xi_{k}(t)+a_{k}^{\dagger}\,e^{-i\vec{k}\cdot\vec{R}}\xi_{k}^{*}(t)\right],
\end{equation}
where the annihilation and creation operators $a_{k}$ and
$a_{k}^{\dagger}$ satisfy the usual commutation algebra
\begin{equation}\label{m5}
[a_{k},a_{k'}^{\dagger}]=\delta ^{(3)}(\vec{k}-\vec{k'}),\quad
[a_{k},a_{k'}]=[a_{k}^{\dagger},a_{k'}^{\dagger}]=0.
\end{equation}
Using the commutation relation (\ref{w11}) and the Fourier
expansion (\ref{m4}), we obtain the normalization condition for
the modes $\xi_{k}(t)$:
\begin{equation}\label{m6}
\xi_{k}(t) \dot{\xi}^*_{k}(t) - \xi^*_{k}(t) \dot{\xi}_{k}(t) = i
\frac{4\pi G }{9}\left(\frac{a_0}{a}\right)^{3},
\end{equation}
where $a(t)=\exp(Ht)$ is the scale factor in (\ref{w1}) and
$a_{0}=\exp(Ht_0)$, $t_{0}$ being the cosmic time at the end of
inflation. Inserting (\ref{m4}) into (\ref{md2}), we obtain that
the modes $\xi_{k}(t)$ satisfy
\begin{equation}\label{m7}
\frac{d^2 \xi_{k}}{d t^2} + 3H\frac{d\xi_{k}}{dt}+ \left[ k^2
e^{-2Ht} +3H^{2}- \lambda^2 \right]\xi_{k} =0.
\end{equation}
By means of Eq. (\ref{m6}) and choosing the Bunch-Davies vacuum
condition, the normalized solution of (\ref{m7}) is
\begin{equation}\label{m8} \xi_{k}(t) =
i\frac{\pi}{3}\sqrt{\frac{G}{H}}\left(\frac{a_0}{a}\right)^{3/2}
{\cal H}^{(2)}_{\nu}[x(t)],
\end{equation}
where ${\cal H}^{(2)}_{\nu}[x(t)]$ is the second-kind Hankel
function, $\nu =[1/(2H)]\sqrt{4\lambda ^2-3H^2}$ and
$x(t)=(k/H)e^{-Ht}$. The parameter $\nu$ is well defined for
values of $\lambda$ that satisfy $\lambda^{2}>(3/4)H$. For
$\lambda =\pm \sqrt{3}H$ the parameter $\nu$ results in the value
$\nu=3/2$, which, as it will be shown in further calculations, is
the value corresponding to a scale-invariant spectrum for the mean
squared scalar fluctuations of the metric $\left<\Psi^2\right>$.
The mean-squared fluctuations for $\Psi$ on the IR sector (on
super Hubble scales), are
\begin{equation}\label{af3}
\left<\Psi^2\right> = \frac{1}{2\pi^2} \int^{\epsilon_0 k_0(t)}_0
dk\  k^2 \xi_{k}
\xi^*_{k}=\frac{2^{2\nu-1}}{9\pi^{2}}(a_0^{3}GH^{2})
\epsilon_{0}^{3-2\nu}\Gamma^{2}(\nu)\left[\frac{\sqrt{\lambda^{2}-(3/4)H^{2}}}{H}\right]^{3-2\nu},
\end{equation}
where $< \,>$ is denoting expectation value, $\epsilon_0\simeq
10^{-3}$ is a dimensionless constant and
$k_{0}(t)=a\sqrt{\lambda^{2}-(3/4)H^{2}}$. Notice that to obtain
the last term of (\ref{af3}), we have employed the asymptotic
expansion ${\cal H}^{(2)}_{\nu}\simeq
(i/\pi)\Gamma(\nu)(x/2)^{-\nu}$ valid for $x\ll 1$. It is easy to
show that the energy density fluctuations associated with
cosmological ($k\ll k_0$) metric fluctuations $\Psi$, are in our
case given by
\begin{equation}\label{af4}
\frac{\delta\rho}{<\rho>}\simeq 2<\Psi^{2}>^{1/2}\simeq 2\Psi.
\end{equation}
Now, employing the equations (\ref{x3}) and (\ref{x14}) and
incorporating the solution (\ref{m2}), we obtain that the 4D
quantum fluctuations for the inflaton field are determined by the
equation
\begin{equation}\label{m9}
\frac{\partial^{2}\delta\Phi}{\partial
t^2}+3H\frac{\partial\delta\Phi}{\partial
t}-e^{-2Ht}\nabla_{R}^{2}\delta\Phi + (6H^2-M^2)\delta\Phi=0,
\end{equation}
where we have used that $[l^{2}(\partial^{2}\delta\varphi/\partial
l^2)+4l(\partial\delta\varphi/\partial
l)]|_{l=H^{-1}}=(M^2/H^2)\delta\Phi$, with $M$ being a separation
constant with mass units. Performing the Fourier expansion of
$\delta\Phi$ in the form
\begin{equation}\label{m10}
\delta\Phi(t,\vec R) = \frac{1}{(2\pi)^{3/2}} {\int} d^3K \left[
a_{K} e^{i\vec{K}\cdot\vec R} \eta_{K}(t) + a^{\dagger}_{K}
e^{-i\vec{K}\cdot\vec R} \eta^*_{K}(t)\right],
\end{equation}
where the annihilation and creation operators $a_{K}$ and
$a_{K}^{\dagger}$ satisfy the commutation algebra given by
(\ref{m5}). Inserting (\ref{m10}) into (\ref{m9}) leads to
\begin{equation}\label{m11}
\frac{\partial^2\eta_{K}}{\partial t^2} + 3 H
\frac{\partial\eta_{K}}{\partial N} + ( K^2 e^{-2Ht} +6H^{2}- M^2)
\eta_{K} =0.
\end{equation}
In this case, due to the commutation relations (\ref{w11}), the
normalization condition reads
\begin{equation}\label{af5}
\dot{\eta}_{K}^{*}\eta_{K}-\dot{\eta}_{K}\eta_{K}^{*}=i\left(\frac{a_0}{a}\right)^{3}.
\end{equation}
Assuming the vacuum Bunch-Davies condition and using (\ref{af5}),
we find that the nor\-ma\-li\-zed solution of (\ref{m11}) can be
written as
\begin{equation}\label{m12}
\eta_{K}(t) =
\frac{i}{2}\sqrt{\frac{\pi}{H}}\left(\frac{a_0}{a}\right)^{3/2}{\cal
H}^{(2)}_{\mu}[y(t)],
\end{equation}
where $\mu= [1/(2H)]\sqrt{4M^{2}-15H^{2}}$
and $y(t) = (K/H) e^{-H t}$. The parameter $\mu$ is well
defined for $M^{2}>(15/4)H^{2}$ and it is always positive in this region.\\

Once we have calculated the normalized modes (\ref{m12}), we are
in position to obtain the mean-squared fluctuations for the 4D
induced inflaton field: $\left<\delta\Phi^{2}\right>$. Following a
similar procedure to that for calculating $\left<\Psi^2\right>$,
we obtain that on super Hubble scales the mean-squared
fluctuations $\left<\delta\Phi^{2}\right>$ are given by
\begin{equation}\label{af6}
\left<\delta\Phi^{2}\right>=\frac{1}{2\pi^2} \int^{\epsilon_1
K_1(t)}_0 dK\  K^2 \eta_{K}
\eta^*_{K}=\frac{2^{-(3-2\mu)}}{\pi^3}\frac{\Gamma^{2}(\mu)}{3-2\mu}\epsilon_{1}^{3-2\mu}
\left[\frac{\sqrt{M^{2}-(15/4)H^2}}{H}\right]^{3-2\mu}a_{0}^{3}H^{2},
\end{equation}
where $K_{1}(t)=a\sqrt{M^{2}-(15/4)H^{2}}$ and we have also used
the asymptotic expansion ${\cal H}^{(2)}_{\mu}\simeq
(i/\pi)\Gamma(\mu)(y/2)^{-\mu}$ valid for $y\ll 1$ to obtain the
last term of (\ref{af3}). From the expression (\ref{af6}) it can
be easily seen that the power spectrum related to the inflaton
quantum fluctuations ${\cal P}_{\delta\Phi}(K)\sim
[K/(aH)]^{3-2\mu}$ is scale invariant when $M=\sqrt{6}H$.
Furthermore, the spectral index given by
$n=4-2\mu=4-\sqrt{3+(4\alpha^{2}-18)}$, with $M=\alpha H$, becomes
$n\simeq 1$ when $\alpha^{2}\simeq 9/2$. In other words, the
spectrum becomes nearly scale invariant around this value of the
dimensionless parameter $\alpha$.\\

\section{Final Remarks}

In this letter we have studied the dynamics of scalar metric
fluctuations from a 5D vacuum state, which is defined on a 5D
background Riemann-flat metric, using ideas of Modern Kaluza-Klein
theory. We passed from a 5D Riemann-flat metric to an effective 4D
metric which describes the de Sitter expansion. The 
dimensional reduction has been achieved by taking the foliation
$l =H^{-1}$ on the metric (\ref{s2}), $H$ being the constant Hubble parameter 
during inflation. From the point of view of a relativistic
observer, this implies that the penta-velocity is null: $U^l=0$.

The most remarkable result here obtained is that, as one would
expect, the spectrum of the squared $\Psi$-fluctuations
$\left<\Psi^2\right>$ on cosmological scales can be scale
invariant, in contrast with the results obtained using a standard
4D formalism\cite{ab,ab1}, where they have a $k^2$-spectrum when
we take a longitudinal gauge. Of course, our results are valid
only on cosmological scales, when vector and tensor perturbations
can be neglected. Another interesting result here obtained is that
the background value for the inflaton field at the beginning of
inflation remains below the Planckian mass:
$\Phi_{b}=\frac{M_{p}}{\sqrt{12\pi}}$ [see Eq. (\ref{m2})].

\section*{Acknowledgements}

\noindent M. A. and M. B. acknowledge CONICET and UNMdP
(Argentina) for financial support and J.E.M.A acknowledges CONACYT
(M\'exico) for financial support.

\end{document}